\documentclass[twocolumn]{aastex631}
\usepackage{amsmath}	
\usepackage{amssymb}	
\usepackage{xcolor}
\usepackage{enumitem}
\usepackage{CJKutf8}

\newcommand{\be}{\begin{equation}}
\newcommand{\ee}{\end{equation}}
\newcommand{\units}[1]{\ensuremath{\, \mathrm{#1}}}  
\newcommand{\Msun}{{\rm M}_\odot}

\begin{document}

\title{Early evidence for isotropic planetary obliquities in young super-Jupiter systems}

\correspondingauthor{Michael Poon}
\email{michael.poon@astro.utoronto.ca}

\author[0000-0001-7739-9767]{Michael Poon}
\affiliation{Department of Astronomy and Astrophysics, University of Toronto, Toronto, Ontario, M5S 3H4, Canada}

\author[0000-0002-6076-5967]{Marta L. Bryan}
\affiliation{Department of Chemical and Physical Sciences, University of Toronto Mississauga, Mississauga, Ontario, L5L 1C6, Canada}
\affiliation{Department of Astronomy and Astrophysics, University of Toronto, Toronto, Ontario, M5S 3H4, Canada}

\author[0000-0003-1927-731X]{Hanno Rein}
\affiliation{Department of Physical and Environmental Sciences, University of Toronto Scarborough, Toronto, Ontario, M1C 1A4, Canada}
\affiliation{Department of Astronomy and Astrophysics, University of Toronto, Toronto, Ontario, M5S 3H4, Canada}

\author[0000-0002-3610-6953]{Jiayin Dong}
\affiliation{Department of Astronomy, University of Illinois at Urbana-Champaign, Urbana, IL 61801, USA}

\author[0000-0003-2573-9832]{Joshua S. Speagle (\begin{CJK*}{UTF8}{gbsn}沈佳士\ignorespacesafterend\end{CJK*})}
\affiliation{Department of Statistical Sciences, University of Toronto, 9th Floor, Ontario Power Building, 700 University Ave, Toronto, ON M5G 1Z5, Canada}
\affiliation{David A. Dunlap Department of Astronomy \& Astrophysics, University of Toronto, 50 St George Street, Toronto, ON M5S 3H4, Canada}
\affiliation{Dunlap Institute for Astronomy \& Astrophysics, University of Toronto, 50 St George Street, Toronto, ON M5S 3H4, Canada}
\affiliation{Data Sciences Institute, University of Toronto, 17th Floor, Ontario Power Building, 700 University Ave, Toronto, ON M5G 1Z5, Canada} 

\author[0000-0002-0924-8403]{Dang Pham}
\affiliation{JILA and Department of Astrophysical and Planetary Sciences, CU Boulder, Boulder, CO 80309, US}
\affiliation{Department of Astronomy and Astrophysics, University of Toronto, Toronto, Ontario, M5S 3H4, Canada}

\begin{abstract}

This decade has seen the first measurements of extrasolar planetary obliquities, characterizing how an exoplanet's spin axis is oriented relative to its orbital axis. These measurements are enabled by combining projected rotational velocities, planetary rotation periods, and astrometric orbits for directly-imaged super-Jupiters. This approach constrains both the spin axis and orbital inclination relative to the line of sight, allowing obliquity measurements for individual systems and offering new insights into their formation. To test whether these super-Jupiters form more like scaled-up planets or scaled-down stars, we develop a hierarchical Bayesian framework to infer their population-level obliquity distribution. Using a single-parameter Fisher distribution, we compare two models: a planet-like formation scenario ($\kappa=5$) predicting moderate alignment, versus a brown dwarf-like formation scenario ($\kappa=0$) predicting isotropic obliquities. Based on a sample of four young super-Jupiter systems, we find early evidence favoring the isotropic case with a Bayes factor of 15, consistent with turbulent fragmentation.

\end{abstract}


\section{Introduction} \label{sec:intro}

Planetary obliquity -- the tilt between a planet’s rotational and orbital axes -- provides a valuable diagnostic of evolutionary history. These obliquities may reflect initial formation mechanisms, or they may be the result of subsequent evolution that can be stochastic (e.g., giant impacts, \citealt{Reinhardt+2020}) or longstanding (e.g., spin-orbit resonances, \citealt{Ward+Hamilton2004}).
While Solar System obliquities have provided valuable case studies of dynamical histories, they remain limited to eight planets sharing a common protoplanetary origin. To move from individual case studies to a statistical understanding, we must look beyond the Solar System to the rapidly expanding catalogue of extrasolar systems, which span a broader range of masses, ages, and orbital architectures.

Only in this decade have \textit{exo}-planetary obliquity constraints become possible \citep{Bryan+2020, Bryan+2021, Palma-Bifani2023, Gandhi+2025, Poon+2024a}, through a rare combination of direct imaging, high-resolution spectroscopy, and space-based photometry for bright, widely-separated ($\gtrsim 50\units{au}$), super-Jupiter systems (e.g., \citealt{Bowler2016}). Four measurements now exist for young ($\lesssim 100$\units{Myr}), massive ($\sim 10-20\units{M_{\rm Jup}}$) companions. These objects lie in the overlap between the upper end of the planetary mass function and the lower end of the substellar mass function \citep{Basri+Brown2006, Chabrier+2014}, offering a window into the formation histories of the outliers of both populations.

Broadly, two formation paradigms are thought to govern in this regime. Giant planets forming bottom-up via core accretion are expected to have initially aligned rotational and orbital axes, inheriting angular momentum from their natal circumstellar disk (e.g., \citealt{Dones+Tremaine1993, Johansen+Lacerda2010, Batygin2018}). In contrast, brown dwarfs forming via gravitational collapse are expected to have randomly-oriented angular momentum vectors \citep{Offner+2016, Lee+2019}. As a result, super-Jupiter obliquities in these two scenarios are expected to differ systematically: aligned or modestly misaligned in the planet-like case, and isotropic in the brown dwarf-like case. 

\begin{figure*}
    \centering
    \includegraphics[width=1.\linewidth]{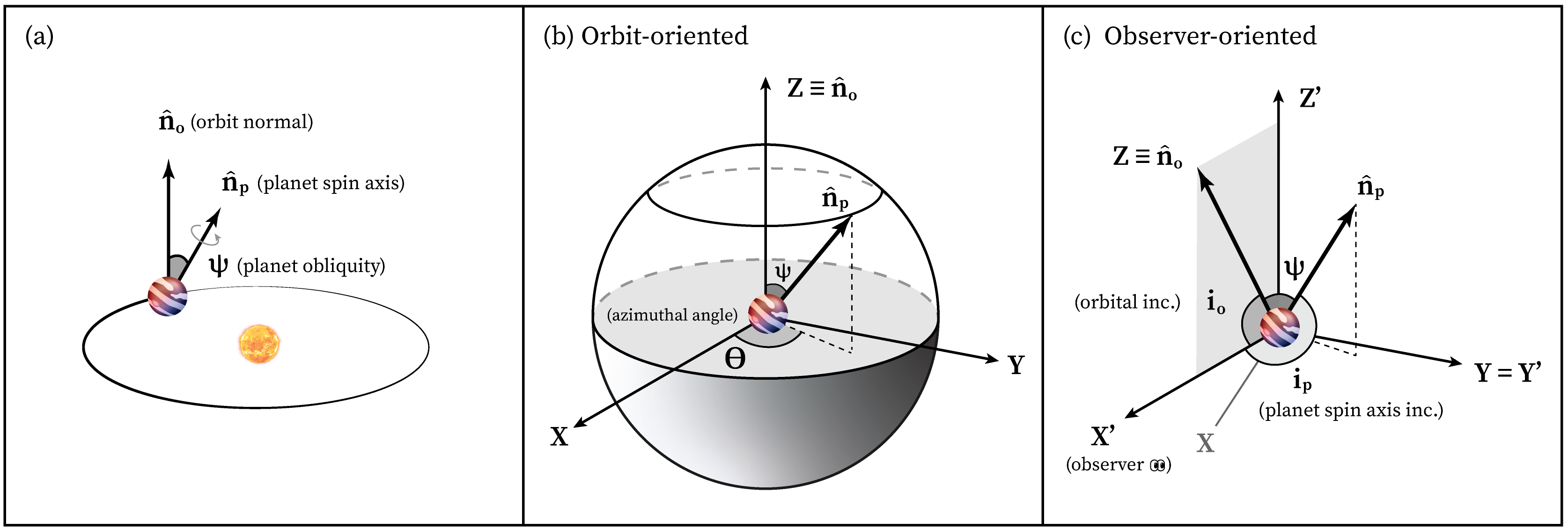}
    \caption{Three illustrations highlighting the relative orientation of a planet spin axis $\mathbf{\hat{n}_p}$ and orbit normal $\mathbf{\hat{n}_o}$: \textbf{(a)} as a cartoon, \textbf{(b)} in an orbit-oriented coordinate system, with the orbit normal along the $Z$-axis, and \textbf{(c)} in an observer-oriented coordinate system, with the observer along the $X'-$axis, and $Y'-Z'$ as the sky plane. Panels \textbf{(b)} and \textbf{(c)} are related by a $90^\circ - i_o$ rotation along the $Y = Y'-$axis. These diagrams were inspired by \citet{Dong+DFM2023}, and adapted from \citet{Poon+2024a}.
    }
    \label{fig:coordinate_system}
\end{figure*}

In this study, we use a hierarchical Bayesian framework to infer the obliquity distribution of four directly-imaged super-Jupiters and test competing formation pathways. By combining individual measurements, we ask: do these systems exhibit a preferred alignment between their spin and orbital axes, or are their spin and orbital axes uncorrelated? Although the current sample is small, it provides an initial test of how obliquities might trace formation pathways, and our analysis presents a framework that can be applied to larger samples in the future. We describe this statistical framework in Section \ref{sec:model} and apply it in Section \ref{sec:application}, with a discussion of observational selection effects. In Section \ref{sec:discussion}, we discuss our results in the broader context of Solar System and stellar binary obliquities. Finally, Section \ref{sec:conclusions} summarizes our conclusions and future prospects for measuring planetary obliquities.

\section{Statistical Framework} \label{sec:model}

Directly-imaged planets lack full 3D spin orientation constraints \citep{Poon+2024a}. As a result, measurements of planetary obliquities $\psi$ are broad for individual systems. Yet across a population, statistical modelling can reveal shared underlying distributions that can point to preferred formation pathways. Ultimately, we want to know: do super-Jupiter obliquities suggest planet-like or brown dwarf-like formation?

A significant challenge to inferring population-level distributions of obliquities is the observational limits on planetary spin measurements. We can constrain a planet's spin axis inclination along the line of sight, $i_p$, by combining measurements of rotational line broadening $v \sin{i_p}$ from high-resolution spectra with independently determined rotation periods from photometric monitoring \citep{Bryan+2018, Zhou+2020b, Masuda+Winn2020}. However, the orientation of the spin axis projected onto the sky plane remains unconstrained. This leaves a key degeneracy in reconstructing the full 3D spin vector $\mathbf{\hat{n}_p}$. In contrast, astrometric monitoring of the planet's orbit provides a direct, 3D measurement of the orbit normal $\mathbf{\hat{n}_o}$ and orbital inclination $i_o$ (e.g., \citealt{Nguyen+2021}). These quantities and their geometric relationship to $\psi$ are illustrated in \autoref{fig:coordinate_system}.

We therefore directly constrain the line-of-sight projected planet obliquity $|i_p - i_o|$, which places a lower bound on the true obliquity angle:
\be
\psi \geq |i_p - i_o|.
\ee
We can de-project $|i_p - i_o|$ using a flexible and informative prior $P(\psi)$, via Bayes' theorem:
\be
P(\psi|i_p, i_o) \propto P(\psi)P(i_p|\psi, i_o), 
\ee
where $P(\psi|i_p, i_o)$ is the posterior of the true obliquity angle, $P(\psi)$ represents the population-level obliquity distribution that serves as the prior for individual systems, and $P(i_p|\psi, i_o)$ is the likelihood. For full details and derivation, see Sec. 3.3 of \citet{Poon+2024a}. 

But what prior shall we choose? Here lies the power of a hierarchical Bayesian framework. It combines noisy obliquity data across a population to uncover the underlying distribution, and the formation story it tells. This insight guides the choice of an appropriate prior, leading to more accurate obliquity measurements for individual systems.

\subsection{Formation Model Setup} \label{sec:model_setup}

\begin{figure}
    \centering
    \includegraphics[width=1.\linewidth]{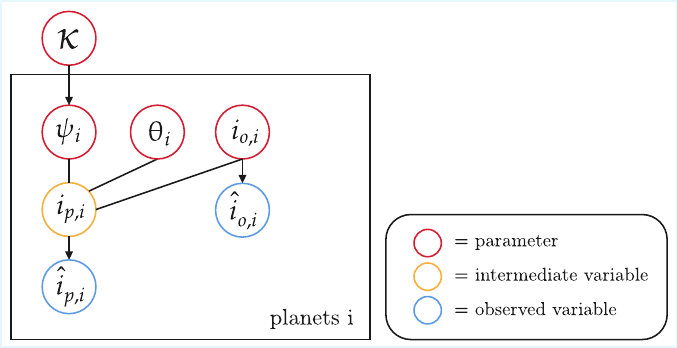}
    \caption{Graphical representation of hierarchical model using quantities depicted in \autoref{fig:coordinate_system}. The arrows represent the direction of data generation from $\kappa$, the population-level obliquity distribution parameter, to $\psi_i$, the obliquity of planet $i$, to the planet's observed variables $\hat{i}_p$ and $\hat{i}_o$.
    }
    \label{fig:plate_diagram}
\end{figure}

We model the population-level distribution of planetary obliquities $P(\psi)$ using a Fisher distribution \citep{Fisher1953}:
\be
P(\psi|\kappa) = \frac{\kappa}{2\sinh{\kappa}}\exp{(\kappa \cos{\psi})}\sin{\psi},
\label{eq:fisher}
\ee
where $\kappa$ is the concentration parameter that sets the mode and spread of the planetary obliquity distribution. This distribution has been widely used in stellar obliquity studies (e.g., \citealt{Fabrycky+Winn2009}, \citealt{Munoz+Perets2018}), as it behaves like a Gaussian on a sphere: small $\kappa$ values produce nearly isotropic orientations, while large $\kappa$ values concentrate $\psi$ near zero. \autoref{fig:plate_diagram} illustrates how $\kappa$ shapes the underlying population and propagates through to the observables.

To avoid biasing our $\kappa$-inference towards either extreme, we adopt a weakly informative hyperprior, following \citet{Munoz+Perets2018}:
\be
P(\kappa) \propto (1 + \kappa^2)^{-1/2},
\label{eq:prior}
\ee
which is approximately uniform when $\kappa \approx 0$, and falls off like $\kappa^{-1}$ for large $\kappa$. This prior is intended to avoid placing strong preference on either perfect alignment or random orientation.

While in the ideal case an abundance of precise measurements would allow us to tightly constrain $\kappa$, our main interest lies in what the population-level obliquity distribution reveals about formation pathways. We therefore define two physically motivated benchmark scenarios.

The central question we aim to answer is whether super-Jupiters retain evidence of their formation pathway in their present-day spin-orbit alignment. Specifically, do they exhibit a broad, isotropic obliquity distribution, consistent with brown dwarf-like formation via top-down fragmentation -- or a more aligned configuration, as might be expected from bottom-up core accretion? To model the former scenario, we adopt a Fisher distribution with concentration parameter $\kappa=0$, yielding $P(\psi) \propto \sin{\psi}$. This corresponds to random orientations between the spin and orbital axes, which are consistent with simulations of gravitational instability in a disk \citep{Jennings+Chiang2021}, and turbulent fragmentation in binaries and multiple-star systems \citep{Offner+2016, Lee+2019}. In our model we define this isotropic case ($\kappa=0$) as the benchmark for brown dwarf-like formation (see upper right panel of \autoref{fig:model_test}).

In contrast, bottom-up formation via core accretion is expected to produce more modest obliquities, as planets inherit their spin from the angular momentum of their protoplanetary disks (e.g., \citealt{Dones+Tremaine1993, Johansen+Lacerda2010, Batygin2018}. To represent this scenario, we define a reference model with $\kappa=5$, producing a Fisher distribution peaked near $\psi \sim 25^\circ$ (see upper left panel of \autoref{fig:model_test}), comparable to the present-day obliquities of Earth, Mars, Saturn, and Neptune, which peculiarly lie in the $20^\circ - 30^\circ$ range. Uranus and Venus have more extreme obliquities near $90^\circ$ and $180^\circ$ respectively, likely due to post-formation evolution, but with only eight Solar System planets, it is unclear whether these are outliers or reflect a broader trend. While our choice of $\kappa=5$ for planet-like formation is heuristic, it captures a representative aligned scenario, with different choices tested in \autoref{appendix:sensitivity_analysis}. Stronger alignment (e.g., $\kappa \gg 5$, such that the peak $\psi \lesssim 1^\circ$) would require much more data to constrain, and is only true for Mercury among Solar System planets. In the following sections, we will describe our framework to infer $\kappa$ from simulated data.

\subsection{Hierarchical Model Setup}
\label{sec:data_gen}

With the formation models in place, we now construct a hierarchical framework that links noisy measurements of each companion's spin axis and orbital inclination to the population-level obliquity distribution, parameterized by $\kappa$. This approach accounts for measurement uncertainties, propagates them through the inference, and tests whether a sample favors low obliquities, consistent with planet-like formation, or a more isotropic distribution, as expected for brown dwarf-like formation.

To motivate the structure of our model, we first describe how simulated data are generated, using the geometry shown in \autoref{fig:coordinate_system} and the conditional dependencies in \autoref{fig:plate_diagram}. We denote the true and observed quantities by $(x, \hat{x})$, and use $\{x_i\}=\{x_i\}_{i=1}^n$ for a population of $n$ independent systems.

Starting from a given $\kappa$, we draw $n$ obliquities $\psi_i$ from a Fisher distribution, each paired with a randomly sampled azimuthal angle $\theta_i \in [0,2\pi]$, and an orbital inclination $i_o$ drawn from an isotropic prior $P(i_o) \propto \sin{i_o}$. From these, we compute the corresponding true spin axis inclinations $i_p$ following Eqn.~(7) of \citet{Dong+DFM2023}. The observed quantities $\hat{i}_o$ and $\hat{i}_p$ are generated by applying observational uncertainties. We use this procedure in Section \ref{sec:model_performance} to validate our hierarchical model with simulated data.

Following this data generation structure and using Bayes' theorem yields the following posterior for the full hierarchical model:
\be
\begin{split}
P\Big(\kappa, \big\{\psi_i, \theta_i, i_{o,i}\big\} \Big| \big\{\hat{i}_{p,i},\hat{i}_{o,i} \big\}\Big) \propto \overbrace{P(\kappa)}^{\text{hyperprior}}\\
\times \underbrace{\prod_{i=1}^n}_{\text{iid}} \underbrace{P\big(\{\psi_i\} | \kappa \big) P\big(\{\theta_i\}\big) P\big(\{i_{o,i}\}\big)}_{\text{prior}}\\
\times \underbrace{P\big(\{\hat{i}_{p,i}\} | \{\psi_i, \theta_i, i_{o,i}\}\big) P\big(\{\hat{i}_{o,i}\} | \{i_{o,i}\}\big)}_{\text{likelihood}},
\label{eq:posterior1}
\end{split}
\ee
where $\kappa, \big\{\psi_i, \theta_i, i_{o,i}\big\}$ are the model parameters, and $\big\{\hat{i}_{p,i},\hat{i}_{o,i}\big\}$ is the data. Since we are ultimately interested in inferring $\kappa$, we collapse our model by integrating out individual system parameters $\big\{\psi_i, \theta_i, i_{o,i}\big\}$. Details of this model collapse and numerical implementation to calculate the final posterior $P\Big(\kappa \Big| \big\{\hat{i}_{p,i},\hat{i}_{o,i}\big\}\Big)$ are provided in \autoref{appendix:model_collapse}.

\subsection{Hierarchical Model Performance} \label{sec:model_performance}

\begin{figure*}
    \centering
    \includegraphics[width=0.85\linewidth]{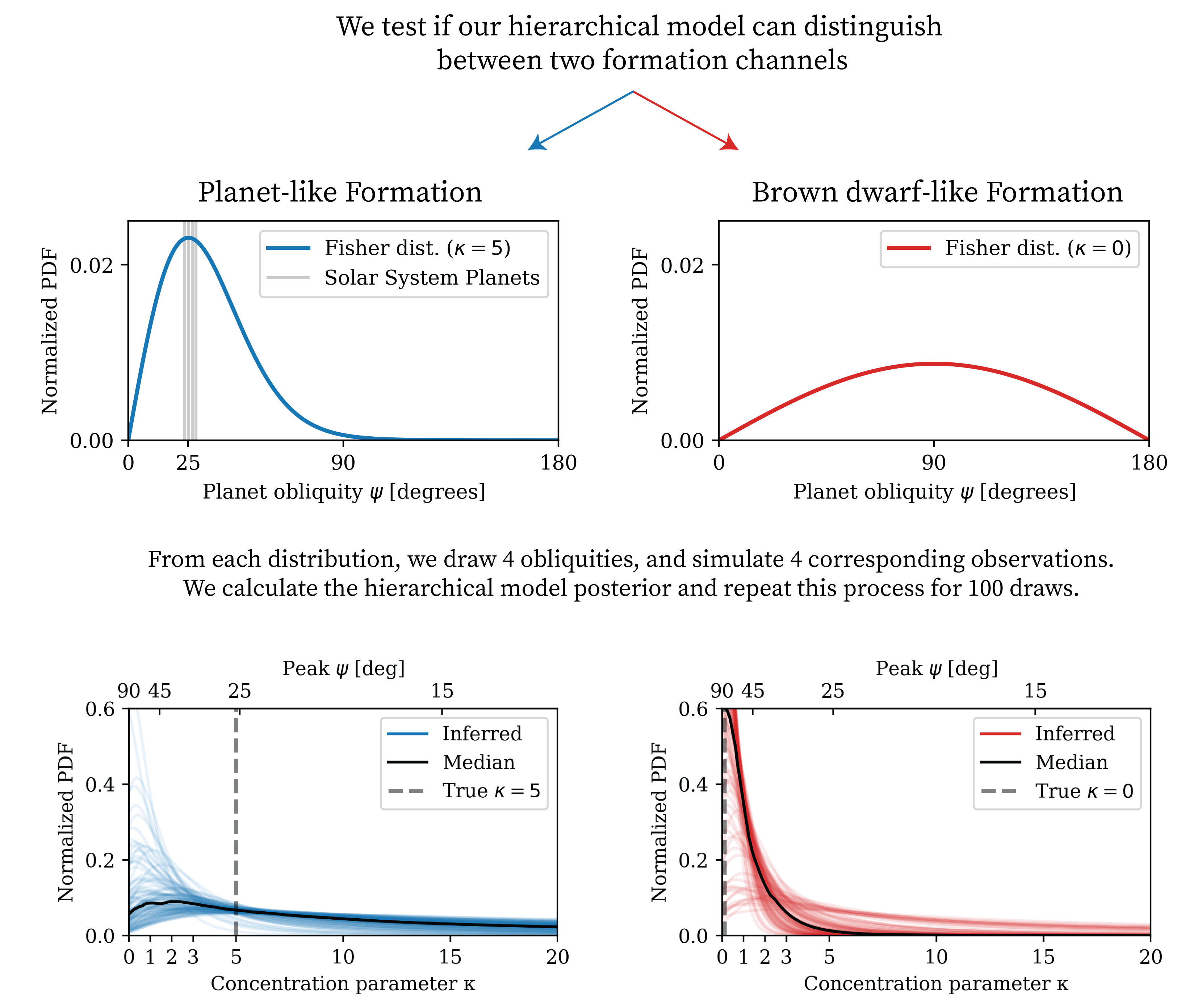}
    \caption{We test whether our hierarchical model can distinguish between planet-like and brown dwarf-like formation using simulated obliquity measurements. \textbf{Top:} Planet-like formation assumes a Fisher distribution with $\kappa=5$ (top-left), while brown dwarf-like formation is modelled as an isotropic distribution with $\kappa=0$ (top-right). Vertical lines mark the obliquities of Earth, Mars, Saturn, and Neptune, which distinctly cluster together. \textbf{Bottom:} For 100 draws of 4 obliquities from the above distribution, we simulate noisy observations for the spin axis and orbital inclinations and infer posteriors on $\kappa$. The median posteriors show qualitatively distinct structure, demonstrating that even 4 measurements can inform formation pathways.
    }
    \label{fig:model_test}
\end{figure*}

\begin{figure*}
    \centering
    \includegraphics[width=0.75\linewidth]{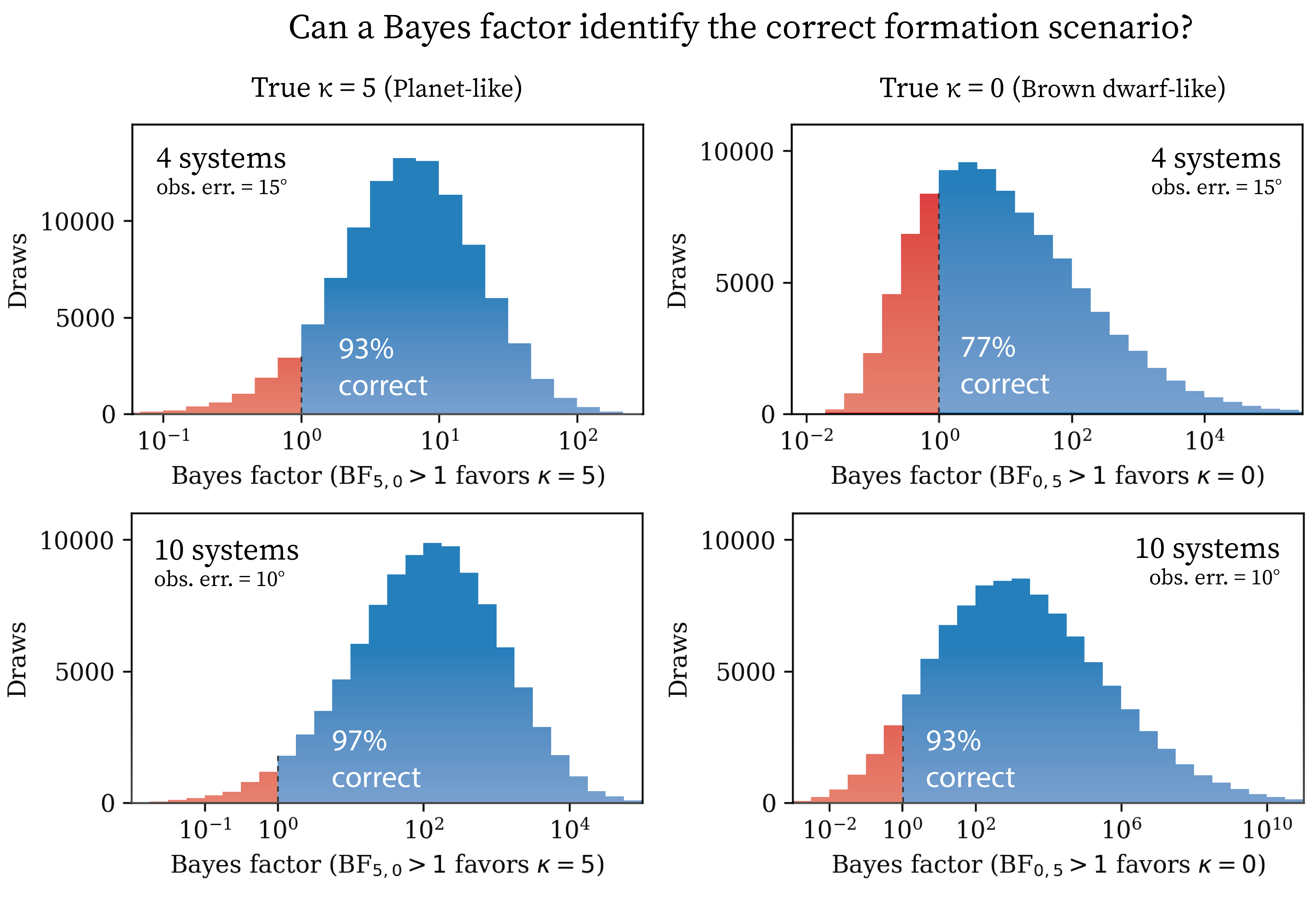}
    \caption{
    We test whether a Bayes factor can correctly identify the underlying formation scenario. For each of 100,000 draws, we simulate observations for 4 or 10 systems. For each draw, we compute the Bayes factor $BF_{5,0}$ if the correct scenario is $\kappa=5$, or $BF_{0,5}$ if the correct scenario is $\kappa=0$. A ratio greater than one favors the correct scenario. \textbf{Top:} With 4 systems and $15^\circ$ uncertainties, the correct scenario is recovered for $93\%$ of the draws for $\kappa=5$ and $77\%$ of the draws for $\kappa=0$. \textbf{Bottom:} Using 10 systems and $10^\circ$ uncertainties improves the accuracy to $97\%$ and $93\%$, respectively.
    }
    \label{fig:model_test_bayes_factor}
\end{figure*}

With the hierarchical model constructed, we test its performance on simulated data, generated from a known true $\kappa$. We conduct two tests: the first test for planet-like formation inference involves drawing 4 obliquities $\psi$ from a Fisher distribution with $\kappa=5$ (upper left panel of \autoref{fig:model_test}), and then generating corresponding spin and orbital inclination $i_p$ and $i_o$ for each system (see Section \ref{sec:data_gen}). We then simulate observational uncertainty by adding Gaussian noise with $\sigma=15^\circ$, to the true values $i_p$ and $i_o$, yielding the observed values $\hat{i}_p$ and $\hat{i}_o$. These uncertainties are typical of existing measurements (e.g. \citealt{Bryan+2020, Bryan+2021, Poon+2024a, Gandhi+2025}). We compute the posterior of our hierarchical model for a single draw (shown as a blue curve in the bottom-left panel of \autoref{fig:model_test}). To capture variability due to small samples, we repeat the process over 100 independent draws, and plot each posterior as a separate curve. The black curve is the median across all draws. 

To test brown dwarf-like formation, we follow the same process as described above but now draw from a true $\kappa=0$ distribution (upper right panel of \autoref{fig:model_test}). The results of our two tests are shown in the bottom panel of \autoref{fig:model_test}, where inference for $\kappa=5$ (blue curves) is qualitatively different than $\kappa=0$ (red curves). While the inference for the brown dwarf-like formation model $\kappa=0$ shows a strong peak near zero, the inference for the planet-like formation model $\kappa=5$ is relatively flat. This is in part because large-$\kappa$ inference is hindered by the size of observational errors, which tend to bias small obliquities to higher values. We investigate the impact of observational uncertainty in a follow-up test described below. Still, the median posteriors exhibit qualitatively distinct structure between the two scenarios, even with just four measurements. In addition, with a small sample size, it is expected that we cannot precisely constrain the value $\kappa$, but that is not required to extract physical insight into formation pathways. While we do test how increasing the sample size improves the $\kappa$ constraint, the more relevant question is which model does the data prefer: $\kappa=0$ or $\kappa=5$?

To answer this, we compute the Bayes factor 
\be
BF_{\kappa_1,\kappa_2} = \frac{P(\{\hat{i}_{o,i}, \hat{i}_{p,i}\}|\kappa=\kappa_1)}{
                P(\{\hat{i}_{o,i}, \hat{i}_{p,i}\}|\kappa=\kappa_2)},
\label{eq:bayes_factor}
\ee
for each draw of $n$ simulated systems, where the marginal likelihoods are evaluated using the collapsed expression given in \autoref{appendix:bayes_factor_calculation}. A Bayes factor $BF_{\kappa_1,\kappa_2}>1$ favors a $\kappa=\kappa_1$ model compared to a $\kappa=\kappa_2$ model. By conventional standards \citep{Jeffreys1998, Kass+Raferty1995}, evidence for $\kappa_1$ is considered weak for values between $1-3$, substantial for $3-10$, strong for $10-100$, and decisive when greater than $100$.

Results are shown in \autoref{fig:model_test_bayes_factor}, using $100,000$ draws that produce a distribution of Bayes factors. In the top panels, planet-like formation is correctly inferred $93\%$ of the time, whereas brown-dwarf formation is correctly inferred $77\%$ of the time. This difference arises because false positives are more common for the brown dwarf-like scenario: it is more likely to sample four small obliquities by chance from an isotropic distribution than to sample a broad range of obliquities from a $\kappa=5$ distribution (top left panel of \autoref{fig:model_test}) where large obliquities are suppressed. We also test the impact of our Gaussian noise assumption by modeling observational errors with a skew-normal distribution (with skew $a=\pm1$). The correct inference rates change by only $\sim 1\%$, indicating a negligible effect.

To assess the impact of improved data quality, we explore two changes independently: increasing the number of systems per draw to 10 and reducing the observational uncertainty to $\sigma=10^\circ$. For the true $\kappa=5$ test, increasing the sample size raises the inference accuracy to $97\%$, while the reduced uncertainty does not affect the inference accuracy. For the true $\kappa=0$ test, increasing the sample size raises the inference accuracy to $90\%$, while reducing the uncertainty yields a more modest improvement to $80\%$. These tests indicate that the most effective way to strengthen the inference is by increasing the number of systems. In the bottom panels of \autoref{fig:model_test_bayes_factor}, we show results for both improvements combined, which may be
attainable in the near future with ongoing observations, such as those enabled by the James Webb Space Telescope.

\section{Application to Exoplanetary Systems} \label{sec:application}

With the hierarchical model evaluated using simulated data, we now turn to the real test: applying it to the handful of planetary-mass companions with available spin-axis and orbital inclinations:

\begin{itemize}[noitemsep]
    \item 2MASS~J01225093–2439505~b (2M0122~b), 
    \item HD~106906~b,
    \item AB~Pictoris~b (AB~Pic~b),
    \item VHS J125601.92-125723.9 (VHS~1256~b).
\end{itemize}

We aim to determine if their population-level obliquity distribution is more consistent with a planet-like or brown dwarf-like formation model ($\kappa=5$ and $\kappa=0$ respectively), as defined in Section \ref{sec:model_setup}. Before presenting the result, we briefly summarize the properties of each system.

\subsection{2M0122~b} 
2M0122~b is a $12-27 \units{M_{\rm Jup}}$ companion at $\sim 52\units{au}$ from its $120\units{Myr}$ $0.4\units{\Msun}$ host \citep{Bowler+2013, Hinkley+2015}. By combining a measured projected rotational velocity $v \sin{i_p} = 13.4 \substack{+1.4 \\ -1.2} \units{km~s^{-1}}$ and orbital inclination $i_o = 103 \substack{+16 \\ -6}^\circ$ with a published rotation period $P_{\rm rot} = 6.0 \substack{+2.6 \\ -1.0}\units{hr}$ from \citet{Zhou+2019}, \citet{Bryan+2020} constrains the first \textit{exo}-planetary obliquity $\psi$, which favors slightly stronger preference for misalignment compared to an isotropic obliquity distribution (i.e., uniform in $\cos{\psi}$).

\subsection{HD~106906~b}
HD~106906~b is a $11.9 \substack{+1.7 \\ -0.8} \units{M_{\rm Jup}}$ companion at $\sim 737\units{au}$ from its $13\units{Myr}$ binary host \citep{Bailey+2014, Nguyen+2021}. \citet{Bryan+2021} combined a measured $v \sin{i_p} = 9.5 \pm 0.2\units{km~s^{-1}}$ with $i_o = 56 \substack{+12 \\ -21}^\circ$ \citep{Nguyen+2021}
and $P_{\rm rot} = 4.1 \pm 0.3\units{hr}$ \citep{Zhou+2020a} to constrain the second exoplanetary obliquity $\psi$, finding evidence for misalignment. Specifically, the posterior is symmetric about $90^\circ$, yielding a mode and 68\% highest probability density interval (HPDI) of $\psi = 55 \substack{+22 \\ -16}^\circ$ for $\psi < 90^\circ$ and $\psi = 125 \substack{+16 \\ -22}^\circ$ for $\psi > 90^\circ$ \citep{Bryan+2021}. This symmetry arises because current observations cannot distinguish whether a planet's spin axis is tilted towards or away from the observer.

\subsection{AB~Pic~b} 
AB~Pic~b is a $10 \pm 1 \units{M_{\rm Jup}}$ companion at $\sim 273\units{au}$ from its $13\units{Myr}$ solar-type (K1V) host \citep{Chauvin+2005, Booth+2021}. \citet{Palma-Bifani2023} combined a measured $v \sin{i_p} = 73 \substack{+11 \\ -27} \units{km~s^{-1}}$, derived from medium-resolution spectroscopy and $i_o = 90 \pm 12^\circ$, with a notably short rotation period $P_{\rm rot} = 2.1$ from \citet{Zhou+2019}, concluding that AB~Pic~b likely has a misaligned planetary obliquity between $45^\circ$ and $135^\circ$. Follow-up work by \citet{Gandhi+2025} measured a much lower $v \sin{i_p} \sim 3.7 \units{km~s^{-1}}$ using high-resolution spectroscopy, and present an updated $i_o = 98 \substack{+12 \\ -5}^\circ$, yet still found a substantially misaligned obliquity.

\subsection{VHS~1256~b} 
VHS~1256~b is a $\sim 12-16 \units{M_{\rm Jup}}$ companion at $\sim 400\units{au}$ from its $140\units{Myr}$ $0.14\units{\Msun}$ binary host \citep{Gauza+2015, Dupuy+2023}. \citet{Poon+2024a} combined a measured $v \sin{i_p} = 8.7 \pm 0.1\units{km~s^{-1}}$ with $i_o = 23 \substack{+10 \\ -13}^\circ$ and $P_{\rm rot} = 22.04 \pm 0.05\units{hr}$ \citep{Zhou+2020b}, finding an obliquity of $\psi = 90^\circ \pm 25^\circ$, reminiscent of Uranus' $98^\circ$ obliquity. Following \citet{Poon+2024a}, we adopt a conservative uncertainty of 10\% ($2.2\units{hr}$) for the rotation period.

\begin{figure*}
    \centering
    \includegraphics[width=0.8\linewidth]{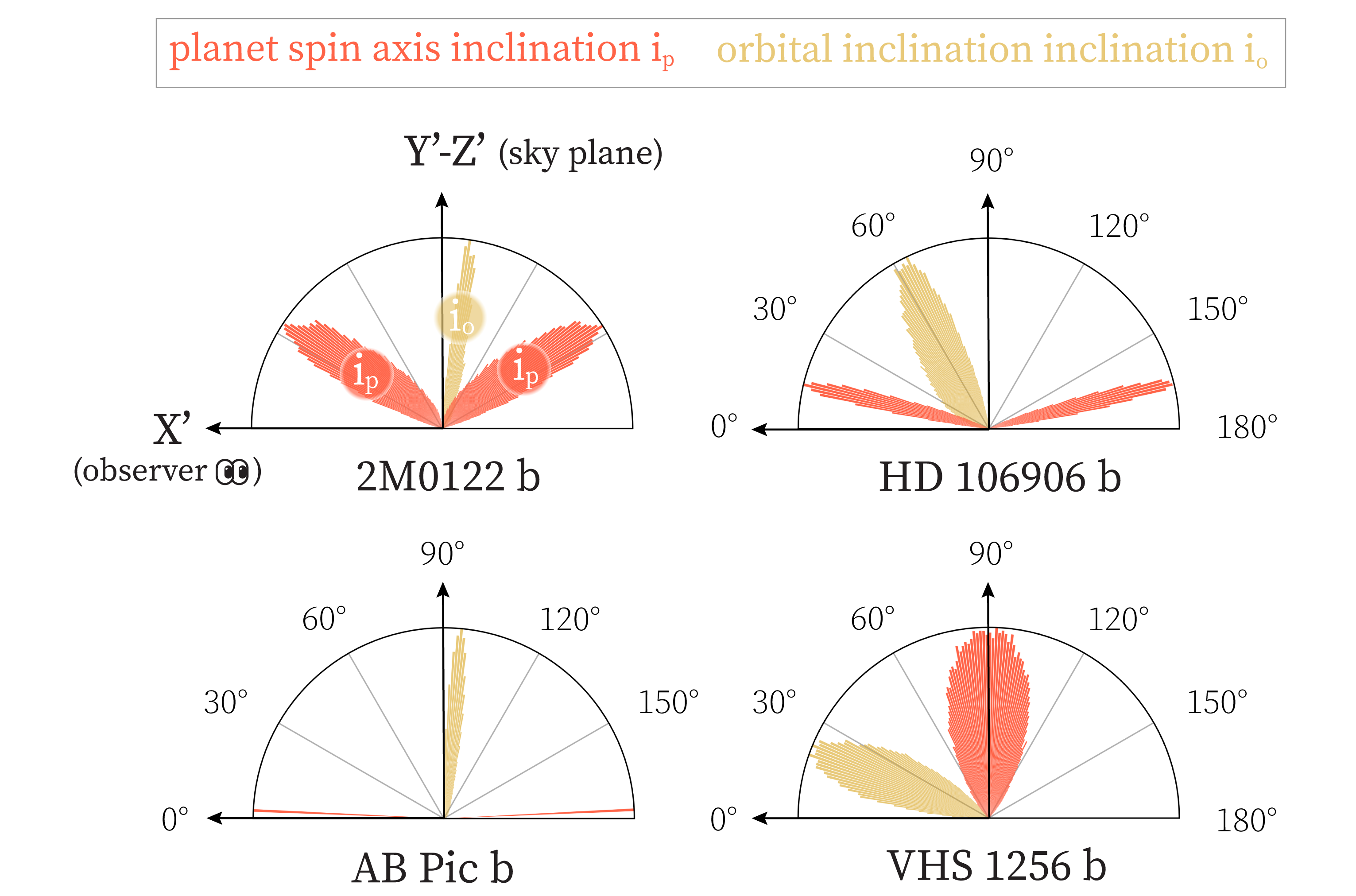}
    \caption{Planet spin axis inclinations ($i_p$, orange) compared with orbital inclinations ($i_o$, gold). These line-of-sight inclinations are relative to an observer along the $X'-$axis (left side), while the vertical axis represents the sky plane ($Y'-Z'$), following \autoref{fig:coordinate_system}. $i_p$ is mirrored about the sky plane due to an observational degeneracy. These diagrams were inspired by \citet{Bowler+2023}.
    }
    \label{fig:polar_plots}
\end{figure*}

The spin-axis and orbital inclinations of these four planetary-mass companions are shown in \autoref{fig:polar_plots}, plotted in orange and gold, respectively. 

\subsection{Early Evidence for Isotropic Obliquities in Young Super-Jupiter Systems} 
\label{sec:early_evidence}

\begin{figure*}
    \centering
    \includegraphics[width=1.0\linewidth]{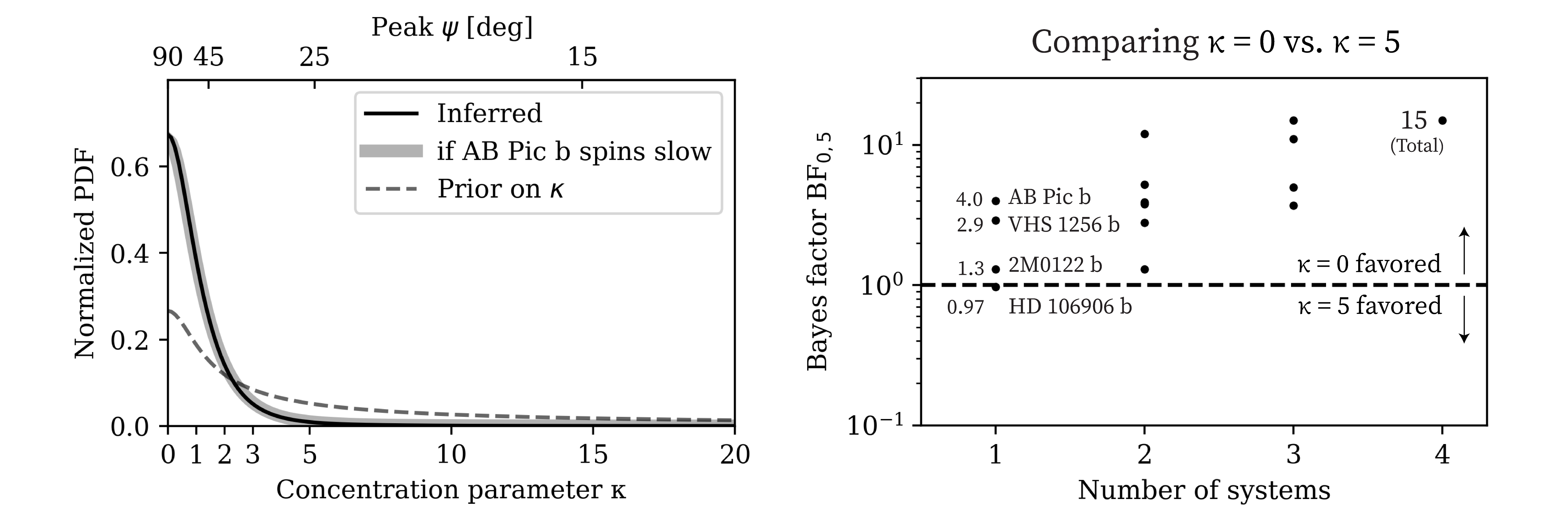}
    \caption{\textbf{Left:} Inferred posterior (\autoref{eq:posterior3}) for the population of four super-Jupiters, using data shown in \autoref{fig:polar_plots}. The shaded gray line assumes a longer rotation period for AB~Pic~b (16.8 \units{hr} instead of 2.1 \units{hr}, in case the observed modulation underestimates the true rotation period (Sec. 3.4.2 of \citealt{Gandhi+2025}). This changes $i_p$ for AB~Pic~b, but does not affect our inference qualitatively. \textbf{Right:} Bayes factor $BF_{0,5}$, where values greater than 1 support $\kappa=0$ over $\kappa=5$, for various subsets of the observed systems. Across all systems, the Bayes factor is 15. This indicates strong but not decisive evidence that super-Jupiter obliquities are consistent with an underlying isotropic distribution, suggestive of brown dwarf-like formation.
    }
    \label{fig:bayes_factor_data}
\end{figure*}

Despite broad uncertainties for individual measurements, our population-level analysis provides early evidence that this sample of super-Jupiters is consistent with an isotropic obliquity distribution. In the left panel of \autoref{fig:bayes_factor_data}, we display the inferred posterior distribution for $\kappa$. The distribution peaks at $\kappa=0$, suggesting that the spin axes of the companions are randomly oriented with respect to their orbital axes -- a configuration broadly consistent with brown dwarf-like formation via gravitational collapse (\citealt{Offner+2016}, \citealt{Lee+2019}).

To quantify the relative preference for a planet-like ($\kappa=5$) vs brown dwarf-like ($\kappa=0$) population-level obliquity distribution, we compute the Bayes factor between the two models, following the model performance tests in Section \ref{sec:model_performance}. Unlike Section \ref{sec:model_performance} where we can simulate an arbitrary number of system draws, the present analysis is limited to a single observed realization of four systems. We find a Bayes factor $BF_{0,5} = 15$, indicating strong support for the $\kappa=0$ model and, by extension, a population of planetary-mass companions consistent with brown dwarf-like formation. In \autoref{appendix:sensitivity_analysis}, we perform a sensitivity analysis to determine how changing our heuristic choice of $\kappa=5$ to represent planet-like formation affects the model comparison, and find that small variations to $\kappa$ or the Fisher distribution still produce a Bayes factor of order 10 favoring the isotropic ($\kappa=0$) scenario.

We show the Bayes factor for different subsets of the observed systems in the right panel of \autoref{fig:bayes_factor_data}. AB~Pic~b and VHS~1256~b contribute the most to the Bayes factor, though further observations are needed to strengthen this conclusion.

We now assess the observational selection effects that may impact our result. For instance, companions with spin axes nearly pole-on to the observer exhibit low-amplitude rotational variability, making it difficult or unfeasible to measure their rotation period \citep{Vos+2020}. This selection effect limits the number of systems amenable to obliquity measurements, but does not bias the inferred obliquity distribution itself, as detectability depends on spin-axis orientation relative to the observer, not orbital inclination. A second selection effect biases against detecting companions on edge-on orbits, since even wide-separation companions mapped via astrometry can pass behind the coronograph at certain projected separations, resulting in periods where they are not observable with direct imaging \citep{Bowler+2023}. While this may restrict obliquity measurements to companions on more face-on orbits, it once again does not bias the obliquities that have been measured.

The limiting factor in our analysis is the small sample size of four planetary obliquity measurements. Despite this restriction, we find evidence favoring an isotropic ($\kappa=0$) obliquity distribution because two of the four systems have large projected obliquities (AB~Pic~b and VHS~1256~b; see \autoref{fig:polar_plots}), which represent lower limits on the true obliquity. The presence of large projected obliquities in two of the four systems is difficult to reconcile with a moderately aligned population ($\kappa=5$), strengthening the preference for an isotropic ($\kappa=0$) obliquity distribution. In the next section, we examine how this emerging picture of isotropic super-Jupiter obliquities compares with trends observed across planetary and stellar systems. 

\section{Discussion: Super-Jupiter Obliquities in Context} \label{sec:discussion}

To place the obliquities of directly-imaged super-Jupiter in context, we compare them with Solar System planets and with stellar binaries. While Solar System planets are broadly expected to exhibit aligned planetary obliquities from bottom-up core accretion, the large tilts of planets such as Uranus and Venus are thought to have been acquired post-formation through evolutionary processes like giant impacts. With six Solar System planets having obliquities under $30^\circ$, the distribution of their obliquities is inconsistent with isotropy \citep{Tremaine1991}, indicating a formation history distinct from super-Jupiters.

For stellar binaries, two broad formation pathways are relevant. In disk fragmentation, companions form within a massive protostellar disk at wide separations ($\gtrsim 10^2\units{au}$), where spin-orbit angles are expected to start roughly aligned. As gas-driven migration shrinks these orbits, accretion from a circumbinary disk can excite both eccentricities and obliquities \citep{BANANA7, Offner2023}. Alternatively, binaries may form through turbulent fragmentation triggered by gravitational instability, in which spin-orbit angles are imprinted at birth with a distribution consistent with isotropy (e.g., \citealt{Offner+2016}).

We next connect these formation pathways with observations of spin-orbit angles in stellar binaries. While wide binaries ($\gtrsim 30\units{au}$) are the closest stellar analogues to super-Jupiters on wide orbits, \citet{Justesen+Albrecht2020} found that their line-of-sight projected obliquities have broad uncertainties and are clustered at $\lesssim 30^\circ$ (see their Fig. 4). Since these are lower bounds, their true obliquities could be much larger, leaving the population consistent with both alignment and misalignment. In contrast, AB~Pic~b and VHS~1256~b have large line-of-sight obliquities (\autoref{fig:polar_plots}) that cannot be deprojected to small values, making them unlikely to be rare draws from a predominately aligned population. While the current sample remains small, these high-obliquity cases suggest that the population of super-Jupiters may exhibit a broad range of spin-orbit angles at formation.

We therefore turn to intermediate-separation ($0.3-2.7\units{au}$) binaries via the BANANA (Binaries Are Not Always Neatly Aligned) survey \citep{BANANA7}, which analyzed 2727 astrometric binaries with F-type primaries using Gaia Data Release 3 \citep{GaiaDR3}. These separations are wide enough to avoid strong tidal realignment, preserving their formation signatures \citep{BANANA7}. Low-eccentricity ($e<0.15$) systems (568 binaries) have small spin-orbit angles, with a mean of $\langle\psi\rangle = 6.9\substack{+5.4 \\ -4.1}^\circ$, whereas for high-eccentricity ($e>0.7$) systems (176 binaries), it rises to $\langle\psi\rangle = 46\substack{+26 \\ -24}^\circ$, indicating systematic misalignment. These trends are consistent with disk fragmentation, with the high-eccentricity systems experiencing dynamical excitation of  eccentricities and obliquities, while the low-eccentricity systems remain largely aligned.

While both eccentric binaries and super-Jupiters show spin-orbit misalignments, their distributions differ: the former appear clustered at intermediate angles, whereas the latter exhibit tentative evidence for isotropic obliquities, consistent with turbulent fragmentation, that establishes spin-orbit misalignments during formation.

To clarify where planetary architectures end and stellar architectures begin, more planetary obliquity measurements are needed in the intermediate regime between $3$ and $50\units{au}$. This separation range is currently unexplored, yet spans the transition from Solar System planets that form via core accretion to wide-orbit companions that resemble brown dwarfs in mass and obliquity. Expanding the sample of obliquities in this range will allow hierarchical models to incorporate a planet-brown dwarf boundary as an inferable population-level feature (e.g., a multi-component model rather than a single $\kappa$ parameter). This would provide a more complete picture of the formation pathways that sculpt planetary and stellar systems alike. 

\section{Conclusions} \label{sec:conclusions}

In this work, we develop a hierarchical Bayesian framework to infer the population-level exoplanet obliquity distribution. Combining four super-Jupiter obliquities within this framework, we find evidence for isotropic obliquities, consistent with brown dwarf-like formation via turbulent fragmentation rather than planet-like formation.

We model the underlying obliquity distribution using a single-parameter ($\kappa$) Fisher distribution, defining a planet-like model ($\kappa=5$, obliquities clustered near $20^\circ-30^\circ$) and a brown dwarf-like model ($\kappa=0$, isotropic obliquities). Our inference favors $\kappa=0$, although we do not report a formal constraint due to the limited number of systems. Comparing the $\kappa=0$ model to the $\kappa=5$ model yields a Bayes factor of 15, with the bulk of support provided by AB~Pic~b and VHS~1256~b, which exhibit large line-of-sight projected obliquities.

Looking forward, the number of planetary obliquity measurements is expected to grow with the James Webb Space Telescope (JWST). Historically, obliquity measurements have been bottlenecked by insufficient photometric precision to detect planetary rotation. JWST's exceptional sensitivity will provide access to planetary rotation period measurements at intermediate separations ($\sim 10\units{au}$), such as the anticipated planetary obliquity measurement of $\beta$~Pictoris~b \citep{Poon+2024b}. Additionally, an independent method to jointly constrain the obliquity and oblateness of transiting planets may be possible with JWST through analysis of transit ingress and egress for rotationally flattened planets (e.g. \citealt{Liu+2024, Lammers+Winn2024, Liu+2025, Price+2025}). At present, current measurements are consistent with planets being spherical, precluding an obliquity constraint, and interpretations are complicated by limb darkening models. Nevertheless, this method has the potential to probe a new and complementary parameter space of close-in, lower-mass planetary obliquities. 

Complementary to photometric approaches, high-resolution spectroscopy is beginning to constrain planetary spin rates ($v \sin{i_p}$) for close-in, directly imaged planets with instruments like HiRISE on the Very Large Telescope \citep{Denis+2025}, while future instruments such as METIS on the Extremely Large Telescope will extend this capability to fainter and cooler systems on Solar System-scale separations.

Beyond planetary obliquities, our hierarchical framework can also be applied to stellar obliquities in non-transiting systems. Future Gaia data releases will expand the catalogue of astrometric orbits for wide-separation planets \citep{Perryman+2014, Wallace+2025}, which is expected to increase the stellar obliquity population significantly. As samples of planetary and stellar obliquities grow, we will better understand how these angles depend on various factors including formation environment, age, host, and binarity. This will bring us closer to a full 3D picture of planetary architecture, encompassing both spin and orbital orientations.

\begin{acknowledgments}

We thank the statistics editor and anonymous reviewer and for their constructive comments that improved this work.
We thank Sam Berek, Joshua Bromley, Mark Dodici, Samantha Fassnacht, and Phil Van-Lane for valuable discussions and feedback that improved this study, and Siddharth Gandhi and Paulina Palma-Bifani for kindly providing the posterior samples for AB Pic b. 
This research has been supported by the Natural Sciences and Engineering Research Council (NSERC) Discovery Grants RGPIN-2023-05173 and RGPIN-2020-04513.
\end{acknowledgments}

\appendix

\section{Sensitivity Analysis} \label{appendix:sensitivity_analysis}

\begin{figure*}
    \centering
    \includegraphics[width=1.0\linewidth]{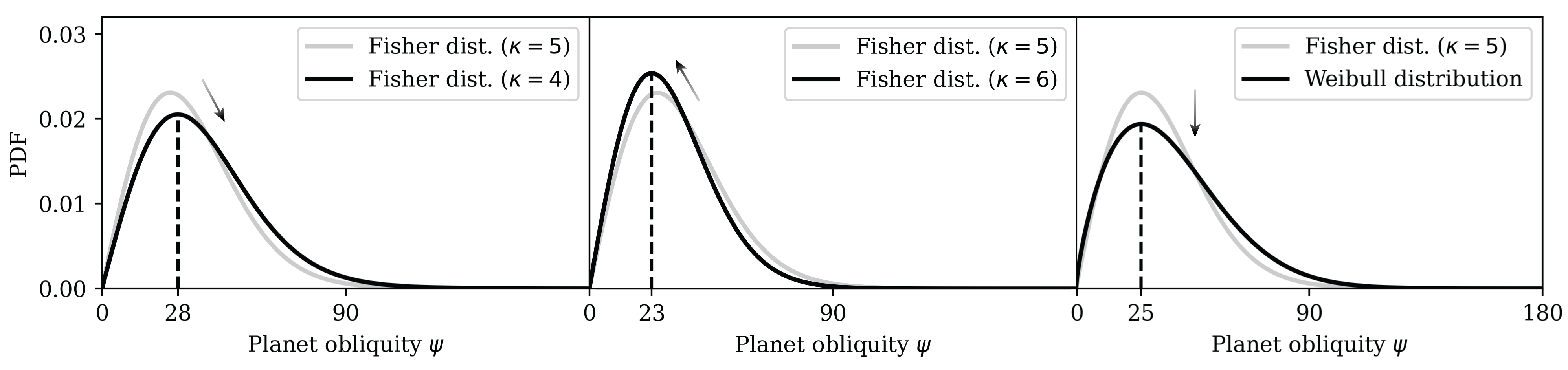}
    \caption{Obliquity distributions for $\kappa=4$, $\kappa=6$, and a Weibull distribution, compared to the fiducial $\kappa=5$ Fisher distribution used to model planet-like formation. The Weibull distribution parameters ($k= 1.75, \lambda=0.71$) match the mode of the $\kappa=5$ distribution, but produce a shallower tail. These alternative choices change the Bayes factor from 15 to 8, 24, and 9, respectively, still favoring a brown dwarf-like ($\kappa=0$) scenario with a Bayes factor of order 10.
    }
    \label{fig:test_prior}
\end{figure*}

\begin{figure*}
    \centering
    \includegraphics[width=1.0\linewidth]{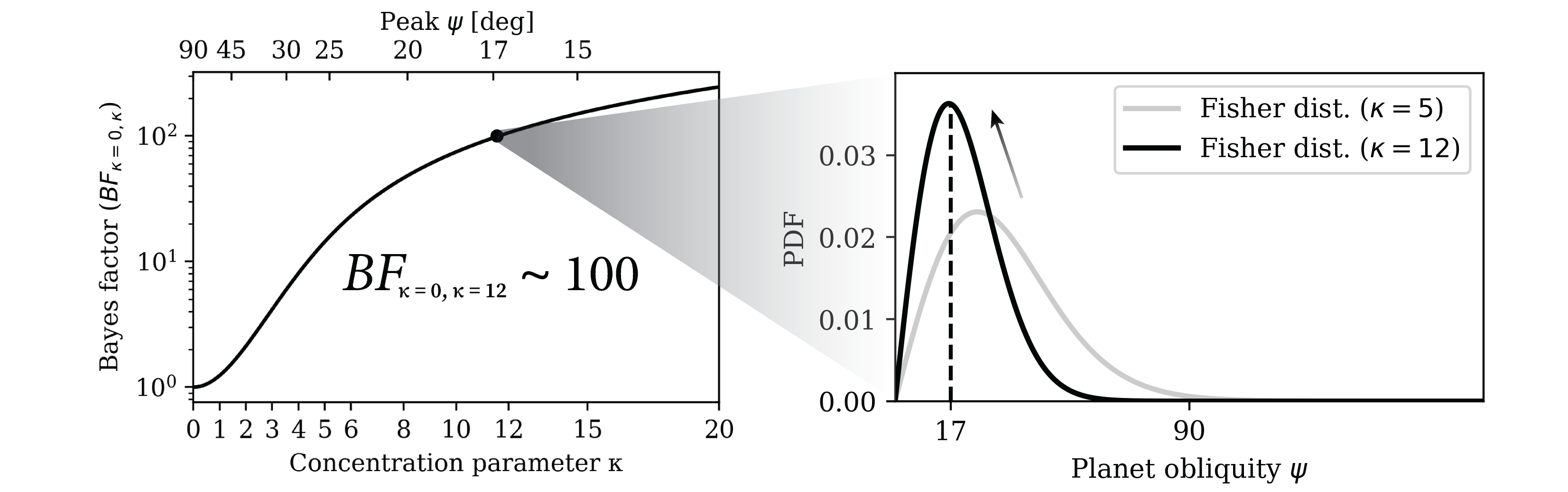}
    \caption{Bayes factors comparing $\kappa=0$ to various $\kappa$. Decisive evidence ($BF \sim 100$) for $\kappa=0$ is reached when the comparison model is  $\kappa \sim 12$, corresponding to an obliquity distribution peaking at $17^\circ$.
    }
    \label{fig:bayes_factors}
\end{figure*}

We test the sensitivity of our results to the assumed prior on the obliquity distribution $P\big(\psi \big| \kappa \big)$. While our main analysis uses a $\kappa=5$ distribution as a fiducial model for planet-like formation, we recompute the Bayes factor using alternative distributions. Specifically, using a $\kappa=4$, $\kappa=6$, or a Weibull distribution (see \autoref{fig:test_prior}), changes the Bayes factor from 15 to 8, 24, and 9, respectively. \autoref{fig:bayes_factors} further shows how the Bayes factor favoring $\kappa=0$ grows with increasing $\kappa$ for the comparison model, reaching decisive evidence ($BF \sim 100$) at $\kappa=12$.
In summary, the isotropic $\kappa=0$ distribution is favored relative to $\kappa=5$ or small variations with a Bayes factor of order 10, and this evidence increases when compared against more concentrated obliquity distributions (larger $\kappa$).

\section{Hierarchical Model Collapse} \label{appendix:model_collapse}

Here, we show how to collapse the hierarchical model to infer only the population-level $\kappa$ parameter. Starting from the full posterior $P\Big(\kappa, \big\{\psi_i, \theta_i, i_{o,i}\big\} \Big| \big\{\hat{i}_{p,i},\hat{i}_{o,i} \big\}\Big)$ in \autoref{eq:posterior1} which contains $3n+1$ parameters (for $n$ systems), we marginalize over the nuisance parameters to obtain the posterior $P\Big(\kappa \Big| \big\{\hat{i}_{p,i},\hat{i}_{o,i}\big\}\Big)$.

We begin by analytically integrating out the azimuthal angle $\theta$, simplifying the model to:
\be
P\Big(\kappa, \big\{\psi_i, i_{o,i}\big\} \Big| \big\{\hat{i}_{p,i},\hat{i}_{o,i} \big\}\Big) \propto \underbrace{P(\kappa)}_{\text{hyperprior}} 
\underbrace{\prod_{i=1}^n}_{\text{iid}} \underbrace{P\big(\psi_i | \kappa \big) P\big(i_{o,i}\big)}_{\text{prior}}  \underbrace{P\big(\hat{i}_{p,i} | \psi_i, i_{o,i}\big) P\big(\hat{i}_{o,i} | i_{o,i}\big)}_{\text{likelihood}},
\label{eq:posterior2}
\ee
We then integrate out $\big\{(\psi,i_o)_i\big\}_{i=1}^n$, yielding:

\begin{align}
P\Big(\kappa \Big| \big\{\hat{i}_{p,i},\hat{i}_{o,i}\big\}\Big)
&= P(\kappa) \prod_{i=1}^n \int P\big(\psi_i | \kappa\big) P(i_{o,i}) P\big(\hat{i}_{p,i} | \psi_i, i_{o,i}\big) P\big(\hat{i}_{o,i} | i_{o,i}\big) \, d\psi_i \, d i_{o,i}.
\end{align}

In practice, we use MCMC chains to sample from the posterior distributions of the true parameters $i_{p,i}$ and $i_{o,i}$, with $m$ samples per system. With these samples, we approximate the integral numerically as an average:
\be
\begin{split}
P\Big(\kappa \Big| \big\{\hat{i}_{p,i},\hat{i}_{o,i}\big\}\Big)
\approx P(\kappa) \prod_{i=1}^n \frac{1}{lm} \sum_{j=1}^m \sum_{k=1}^l P\big(\psi_i^{(k)} \big| \kappa \big) P\big(\hat{i}_{p,i} \big| \psi_i^{(k)}, i_{o,i}^{(j)}\big),
\label{eq:posterior3}
\end{split}
\ee
where we have divided out $P\big(\hat{i}_{o,i} \big| i_{o,i}\big) P\big(i_{o,i}\big)$ so that the posterior is sampled uniformly in $i_o$. Here, $P\big(\kappa \big)$ is given in \autoref{eq:prior}, and the Fisher distribution $P\big(\psi \big| \kappa\big)$ is given in \autoref{eq:fisher}. To compute $P\big(i_p \big|\psi,  i_o \big)$, we use Eqn.~(11) of \citet{Poon+2024a}.

\section{Bayes Factor Calculation} \label{appendix:bayes_factor_calculation}

To compare between models using the Bayes factor, as in Sections \ref{sec:model_performance} and \ref{sec:early_evidence}, we use the marginal likelihood or evidence of a model given a $\kappa=\kappa_1$:

\begin{align}
P\Big( \big\{\hat{i}_{p,i},\hat{i}_{o,i}\big\} \Big| \kappa=\kappa_1 \Big)
&= \prod_{i=1}^n \int P\big(\psi_i | \kappa=\kappa_1\big) P(i_{o,i}) P\big(\hat{i}_{p,i} | \psi_i, i_{o,i}\big) P\big(\hat{i}_{o,i} | i_{o,i}\big) \, d\psi_i \, d i_{o,i},
\end{align}

where we assume that all $n$ systems are independent and identically distributed.

\bibliography{main}{}
\bibliographystyle{aasjournal}

\end{document}